\documentclass[fleqn,twoside]{article}
\usepackage{espcrc2}
\usepackage{amssymb}

\usepackage{graphicx}
\usepackage[figuresright]{rotating}

\newcommand{\AmS}{{\protect\the\textfont2
  A\kern-.1667em\lower.5ex\hbox{M}\kern-.125emS}}

\title{MAGNETIZATION OF CHARGE-ORDERED La$_{2-x}$Sr$_x$NiO$_{4+\delta}$}

\author{P. G. Freeman\address[MCSD]{Department of Physics, Oxford University,
 Oxford, OX1 3PU, United Kingdom}%
        \thanks{coressponding author. Tel: +44 1865 272308;
                fax; +44 1865 272400
                E-mail address: p.freeman1@physiscs.ox.ac.uk
                URL;
                http://xray.physics.ox.ac.uk/Boothroyd/Default.htm  \newline
                We would like to thank the Engineering and Physical Sciences Research
                Council of Great Britain for their help in funding this work, and the
                support received from the Wohlfarth Fund. },
        A. T. Boothroyd\addressmark,
        D. Prabhakaran\addressmark,
        D. Gonz\'{a}lez\address{Instituto de Ciencia de Materials de Arag\'{o}n, CSIC-Universidad de Zaragoza, 50009 Zaragoza,
Spain}}

\begin{document}

\begin{abstract}
We report magnetization measurements on
La$_{2-x}$Sr$_x$NiO$_{4+\delta}$ single crystals, with $0 \leq x
\leq 0.5$. Glassy behaviour associated with the formation of
spin-charge stripes, and a separate spin-glass phase at low
temperatures were observed. We have also found a `memory effect'
in the magnetic field -- temperature history, which is found to be
suppressed in the low temperature spin state of the $x=0.33$
crystal. \vspace{1pc}
\end{abstract}

%\begin{keyword}
%$\sep$ charge stripes$\sep$charge ordering,La2-xSrxNiO4

%\PACS code  $\sep$ 71.45.Lr $\sep$ 75.30.Fv $\sep$ 75.50.Lk
%\end{keyword}

%\end{frontmatter}

\maketitle

Spin and charge ordering in the form of stripes has been observed
in a number of doped antiferromagnets, particularly in layered
transition-metal oxides such as cuprates and
nickelates\cite{tranquada-Nature-1995}. The possibility that
stripe correlations play an important role in the mechanism of
high-$T_{\rm c}$ superconductivity makes it important to
understand the properties of the stripe phase.

In La$_{2-x}$Sr$_x$NiO$_{4+\delta}$ (LSNO) the charges order into
periodically spaced lines of charges that lie at 45$^{\circ}$ to
the Ni-O bonds\cite{chen-PRL-1993}. At lower temperatures the
Ni$^{2+}$ spins order antiferromagnetically between the
charge-stripes that act as antiphase domain walls to the magnetic
order\cite{yoshizawa-PRB-2000}. These studies have also revealed
commensurability effects occurring only in the 1/3 and 1/2 doped
materials\cite{lee-PRB-2001,me}. Previous work has also found a
spin glass state exists at low temperatures in
LSNO\cite{spinglass}. Here we report magnetization measurements
that reveal other types of glassy behaviour in these materials,
including a field `memory effect' in the spin glass state.

\begin{figure}[htb]
\includegraphics*{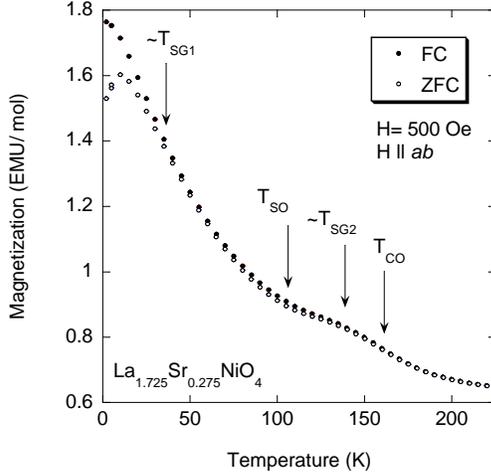}|
%\vspace{9pt}
\caption{\small A typical set of ZFC and FC data for a LSNO
material, showing the convergence of the FC and ZFC magnetization
between $T_{\rm SO}$ and $T_{\rm CO}$. The charge ordering,
$T_{\rm CO}$, (as confirmed by x-ray scattering\cite{Hatton}) spin
ordering $T_{\rm SO}$, (as confirmed from neutron diffraction) and
spin glass temperatures, $T_{\rm SG1}$ and $T_{\rm SG2}$, are
indicated.} \label{fig:typical}
\end{figure}

\begin{figure}[htb]
\includegraphics*{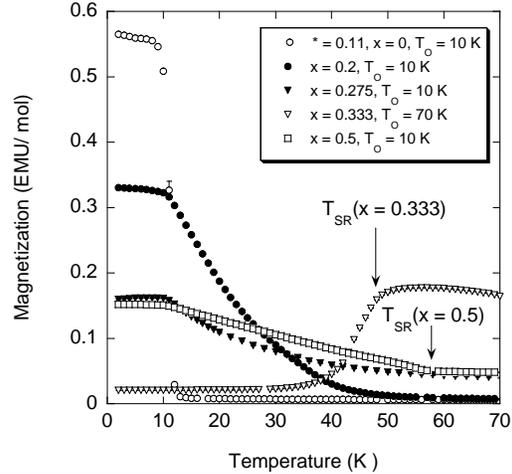}|
%\vspace{9pt}
\caption{\small The signal induced by field-cooling to the
temperature $T_{\rm 0}$ indicated in the key, then cooling to 2 K
in zero field, and measuring while warming in zero field. The
spin-reorientation temperatures of the $x=0.333$ and $x=0.5$
samples are indicated. } \label{fig:inducedFC}
\end{figure}

We performed magnetization measurements on single crystals of
La$_{2-x}$Sr$_x$NiO$_{4+\delta}$ using a SQUID magnetometer
(Quantum Design). The single crystals were grown by the
floating-zone technique\cite{Prab} from high purity oxides. These
crystals have a typical size of $5 \times 5 \times 2\ $mm. The
oxygen content was determined by thermogravimetric analysis.

Figure \ref{fig:typical} shows a set of typical results, obtained
from a sample with $x$ = 0.275 with a measuring field of 500 Oe
applied parallel to the $ab$ plane. The charge and spin ordering
temperatures are indicated. A spin glass phase at low temperature
($T_{\rm SG1}$ $\approx$ 40 K) is signalled by a clear divergence
of the zero-field-cooled (ZFC) and field-cooled (FC)
magnetization. However, glassy behaviour is seen to extend to much
higher temperatures: a smaller ZFC/FC splitting only vanishes at a
temperature $T_{\rm SG2}$ between $T_{\rm SO}$ and $T_{\rm CO}$.
In the case of $x$ = 0.5, where $T_{\rm SO}$ $\sim$ 90 K and
$T_{\rm CO}$ $\sim$ 480 K, the ZFC and FC curves did not close up
until well above T$_{SO}$. Hence the $T_{\rm SG2}$ feature appears
to be correlated with charge-ordering.

In the spin glass phase of LSNO we observed all the materials to
have an effective `memory' of their temperature-field history in
the $ab$ plane. We observed this by cooling the sample in a field
of 500 Oe to $T_{\rm 0}$ ($<$ $T_{\rm SG1}$), removing the field,
cooling to 2 K, and then measuring in zero field while warming.
Apart from the $x$ = 0.333 sample, we observe a stable ($>$ 1 hr)
induced magnetic signal that decreased in size above $T_{\rm 0}$,
and finally levels off to a constant value at $T_{\rm SG1}$, see
figure \ref{fig:inducedFC}. With $x$ = 0.333, the signal is small
at low temperatures, but rises sharply just below $T_{\rm SR}
\approx$ 50 K, corresponding to a known spin
reorientation\cite{lee-PRB-2001}. The $x$ = 0.5 material also has
a spin reorientation\cite{me}, but no suppression of this
FC-induced signal was observed in this material, although a change
in the slope can be seen at $T_{\rm SR} \approx$ 57 K, which is
$\simeq$ $T_{\rm SG1}$ in this material. Again this shows how the
commensurability  of the 0.333 and 0.5 doped materials cause them
to behave differently from other doping levels.

Our results further add to the knowledge of the glassy behaviour
associated with charge-stripe ordering in LSNO. The existence of
two apparently different glassy regimes, as well as other unusual
field-temperature hysteretic effects, shows that the glassy
properties of stripe phases are quite different to those observed
in many other spin-glass materials. Further work needs to be
carried out to determine the nature of these effect in fields
parallel to $c$.

\end{document}